\documentclass[preprint,authoryear,12pt]{elsarticle}

\usepackage{graphics}
\usepackage{float}

\usepackage{comment}

\usepackage{float}
\usepackage{graphicx}
\usepackage{wrapfig}

\usepackage[draft]{todonotes}   

\usepackage[labelfont=bf]{caption}
\usepackage{subcaption}
\usepackage{color}
\usepackage{amsmath} 
\usepackage{mathtools} 
\usepackage{amsfonts} 
\usepackage{amssymb}

\usepackage{array}
\usepackage{booktabs}
\usepackage{rotating}
\usepackage{multirow}
\usepackage{multicol}
\usepackage{lscape}

\usepackage[export]{adjustbox}

\usepackage{color}

\usepackage[draft]{todonotes}   
 
\usepackage{soul}

\definecolor{darkgreen}{rgb}{0.53, 0.66, 0.42}

\usepackage[colorlinks=true,linkcolor=blue,citecolor=blue]{hyperref}

\usepackage[ruled,vlined]{algorithm2e}

\usepackage[normalem]{ulem} 

\journal{arXiv (preliminary version)}

\begin{document}

\begin{frontmatter}



\title{Population Template-Based Brain Graph Augmentation for Improving One-Shot Learning Classification}

\author{Oben \"{O}zg\"{u}r\fnref{BASIRA}}
\author{Arwa Rekik\fnref{MED}}

\author{Islem Rekik\corref{cor}\fnref{BASIRA,IMPERIAL}}

\address[BASIRA]{BASIRA lab, Faculty of Computer and Informatics, Istanbul Technical University, Istanbul, Turkey}
\address[MED]{Faculty of Medicine of Sousse, Tunisia \ }

\address[IMPERIAL]{Imperial-X Computing, Imperial College London, UK \ }

\cortext[cor]{Corresponding author; Dr Islem Rekik, \url{http://basira-lab.com/}, i.rekik@imperial.ac.uk}


\begin{abstract}

\textbf{-Background.} 
The challenges of collecting medical data on neurological disorder diagnosis problems paved the way for learning methods with scarce number of samples. Due to this reason, one-shot learning still remains one of the most challenging and trending concepts of deep learning as it proposes to simulate the human-like learning approach in classification problems. 

\textbf{-Previous Work.}
Previous studies have focused on generating more accurate fingerprints of the population using graph neural networks (GNNs) with connectomic brain graph data. Thereby, generated population fingerprints named connectional brain template (CBTs) enabled detecting discriminative bio-markers of the population on classification tasks. However, the reverse problem of data augmentation from single graph data representing brain connectivity has never been tackled before.

\textbf{-Proposed Method.} 
In this paper, we propose an augmentation pipeline in order to provide improved metrics on our binary classification problem. Divergently from the previous studies, we examine augmentation from a single population template by utilizing graph-based generative adversarial network (gGAN) architecture for a classification problem.

\textbf{-Results.} 
We benchmarked our proposed solution on AD/LMCI dataset consisting of brain connectomes with Alzheimer's Disease (AD) and Late Mild Cognitive Impairment (LMCI). In order to evaluate our model's generalizability, we used cross-validation strategy and randomly sampled the folds multiple times.

\textbf{-Conclusion.}
Our results on classification not only  provided  better accuracy when augmented data generated from one sample is introduced, but yields more balanced results on other metrics as well. 

\end{abstract}

\begin{keyword}
One-shot Learning \sep Connectional Brain Template \sep Graph Neural Network \sep graph-based Generative Adversarial Network 
\end{keyword}

\end{frontmatter}

\section{Introduction}


In the recent years, deep learning methodologies have achieved great success regarding classification problems of neurological disorders \citep{Li:2015}. Despite such progress, conventional deep learning models fail to learn efficiently from connectomic brain graph data since conventional methods are incompetent to capture the topology of non-Euclidean geometries \citep{cao2022geometric}. Fortunately, GNNs have provided learning by considering the topology and connectivity of regions of interests (ROIs) inside the brain which extracts a deeper interpretation of the brain networks \citep{Song:2020}. Thereby, GNNs and brain connectome data have become one of the most reliable and vigorous components in the domain of network neuroscience \citep{Bessadok:2021}. Despite these enhancements, what most deep learning based models suffer from commonly is the scarcity of data since gathering medical data is time-taking and laborious due to privacy concerns. These circumstances pave the way for boosting the learning methods conditioned by a limited number of samples such as one-shot learning with medical data. Such concept is challenging due to its nature since deep models tend to overfit when they are trained with small data. To overcome this hassle, one common approach is to introduce data augmentation to increase the generalizability of the model. Although recent studies have focused on data augmentation techniques on different domains and different data types such as images and texts \citep{Zhou:2018, Chen:2019, Gao:2018}, graphs bring an extra layer of challenge to the problem of data augmentation due to their complex and non-Euclidean structure explaining the current few studies focusing specifically on graph data.

In this paper, we adapt a very unusual methodology of learning and augmenting based on a single population template, namely a CBT for the case of connectomic brain graph data. Recent studies have focused on generating a well-centered brain network atlas of the population which also enables to discern discriminative bio-markers of the brains between different populations in a more peculiar way \citep{Rekik:2017}, and thereby motivated us to question the capability of population-templates on learning tasks. Even though many different networks have been proposed to generate population-templates by fusing multi-view connectomic brain graph data \citep{SNF,DGN,MGN}, very few studies have considered the idea of using the CBT on a one-shot learning diagnostic task \citep{Guvercin:2021}. Moreover, such studies ruled out the challenge of augmentation using only the provided population-template to achieve a better performance. To tackle this intimidating problem, we are inspired to utilize the gGAN architecture originally used for graph normalization with respect to a fixed CBT to solve the augmentation problem from one single CBT \citep{gGAN}. 

Few studies have focused on the problem of data augmentation from graphs specifically for graph classification. \citep{Zhou:2020} have proposed M-Evolve framework for graph augmentation. M-Evolve's graph augmentation method relies on random mapping and motif-similarity mapping algorithms which are heuristic, thereby the algorithms heuristically modify the edges of the graphs taken from a candidate set. Even though M-Evolve produces robust benchmarks on various datasets, it does not study any deep graph paradigm to constrain the generated samples' variability. Another study \citep{Han:2022} have adapted the mixup technique and introduced $\mathcal{G}$-Mixup to improve graph classification robustness on GNNs. $\mathcal{G}$-Mixup makes use of graphons of a specific class as generators. Then, $\mathcal{G}$-Mixup interpolates the features and labels of two graphs that are selected randomly. $\mathcal{G}$-Mixup have provided better performance and robustness on different datasets by interpolating graphons of different graph classes for the first time. However, the generated samples are biased by the two samples selected from training data. \citep{Graa:2019} proposed a Multi-View LEArning-based data Proliferator (MV-LEAP) for the context of brain graph proliferation. However, MV-LEAP is specifically designed for highly-imbalanced classes with multiple views. Such works and limitations have motivated us to ask the the following question: \textit{Given a
CBT that catches the variabilities and patterns of the population in a better way, can we augment brain graph data to improve one-shot classification performance between brain graphs?}

To address those limitations and to seek an answer to our question, we considered the idea of using a single CBT while generating synthetic brain graphs. Moreover, we used a graph based generative adversarial network architecture named gGAN to learn from the graph topology. Consequently, we constrained the variability and patterns of the generated samples by a well-centered and discriminative population template. Firstly, we train the graph based augmentator network containing a generator and a discriminator by only feeding one single CBT for each of the populations independently. Second, we generate symmetric random noise matrices representing brain connectivity. By one forward propagation of random noise matrix on the trained generator, we get the synthetic brain graphs to include in our training set. Finally, we classified the brain graphs using random forest classifier and compared our baseline results with one-shot learning.

Explicitly, we proposed an augmentator network to generate synthetic brain graphs by only learning from the population template. To the best of our knowledge, the idea of graph augmentation from a single population has never been investigated in the domain of neuroscience network. Consequently, our work presents the following contributions to the field: (i) It tackles the dreadful problem of augmenting from one single sample. (ii) It introduces the idea of augmenting using a population template to catch the variability in a more sophisticated way. (iii) We utilize a deep generative adversarial graph architecture rather than relying on heuristic algorithms.


\section{Proposed method}


\textbf{Problem statement.}
Given two CBTs drawn from different populations (e.g., Alzheimer's Disease and Late Mild Cognitive Impairment) representing edge connectivity between ROIs of size $n_r\times n_r$, CBTs are denoted as ${{X}}_{1,c1}^{CBT}$ and ${{X}}_{1,c2}^{CBT}$. Our aim is to design a pipeline for class-based data augmentation by implementing a graph-based generative adversarial network paradigm and augmenting only on a single population template to boost one-shot graph classification task performance. To generate synthetic brain graph data, we feed a set of symmetric random noise matrices of size $n_r\times n_r$ representing edge connectivity denoted as $\{{{X}}_{1}^{R}, {{X}}_{2}^{R}, {{X}}_{3}^{R}, ..., {{X}}_{n}^{R} \}$ with $n$ being the desired number of augmented samples for one class.  

\begin{figure}[H]
	\centering
	\vspace{-10pt}
	{\includegraphics[width=14.5cm]{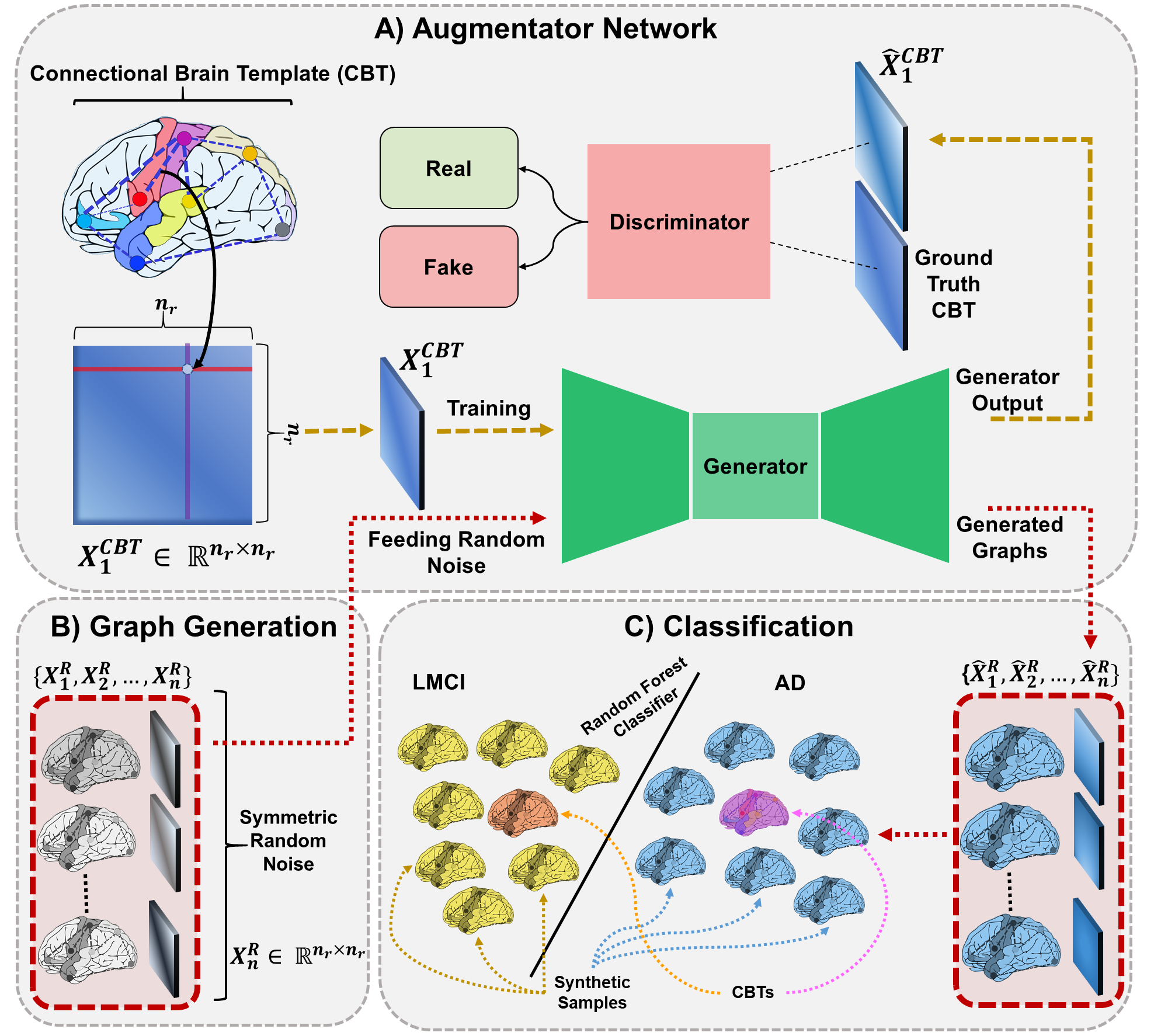}}
	\caption{\scriptsize \emph{Illustration of the proposed network to generate synthetic brain graphs from population template.} (A) Augmentator Network. Ground truth CBT is denoted with ${{X}}_{1}^{CBT}$. The generator is trained by only using a single sample which is the CBT whereas the discriminator tries to distinguish the generated $\hat{{X}}_{1}^{CBT}$. For each class the network is trained independently. (B) Graph Generation. Symmetric random noise matrices representing brain connectivity are denoted in set $\{{{X}}_{1}^{R}, {{X}}_{2}^{R}, {{X}}_{3}^{R}, ..., {{X}}_{n}^{R} \}$. Each random noise matrix is fed into the generator to generate the synthetic brain graphs for both classes. (C) Classification. Generated brain graphs are denoted with $\{\hat{{X}}_{1}^{R}, \hat{{X}}_{2}^{R}, \hat{{X}}_{3}^{R}, ..., \hat{{X}}_{n}^{R} \}$ Finally, each brain graph's upper-triangular part is vectorized and classified using random forest classifier.} 
	\label{fig:main}
\end{figure}

We exhibited each key step and component of the proposed augmentation pipeline. Figure~\ref{fig:main} describes each of the three steps of the pipeline: (a) Augmentator network, (b) Graph Generation, and (c) Classification. Table~\ref{tab:1} represents displays of all mathematical notations used throughout the paper.

\renewcommand{\arraystretch}{1.4}

\begin{table}[H]
	\caption{ \emph{Mathematical notations used in the paper.}\label{tab:1}}
	\begin{scriptsize}
		\begin{tabular}{c@{~~}c@{~~}c}
			\toprule
			Mathematical notation & Definition \\
			\midrule
			
            $n_r$ & number of regions of interest (ROIs) or nodes in a connectome\\
            
            $n$ & number of generated samples\\
            
            $G$ & GAN Generator\\
            
            $D$ & GAN Discriminator\\
            
            $\mathcal{L}_{comp}$ & complete loss function\\
            
            $\mathcal{L}_{adv}$ & adversarial loss function\\
            
            $\mathcal{L}_{L1}$ & $L_1$ loss function\\
            
            $\lambda$ & $L_1$ loss parameter\\
            
            ${\mathbf{X}}_{1}^{CBT}$ & ground truth connectional brain template (CBT)\\
            
            $\hat{\mathbf{X}}_{1}^{CBT}$ & generator output while training\\
            
            ${\mathbf{X}}_{n}^{R}$ & $n$-th random noise graph\\
            
            $\hat{\mathbf{X}}_{n}^{R}$ & $n$-th synthetic graph\\
            
            ${{\mathbf{X}}}_{1,c1}^{CBT}$ & CBT of class 1\\
            
            ${{\mathbf{X}}}_{1,c2}^{CBT}$ & CBT of class 2\\
            
            ${\mathbf{X}}_{c1}^{tr} = \{ {{\mathbf{X}}}_{1,c1}^{CBT}, \hat{\mathbf{X}}_{1,c1}^{R}, \hat{\mathbf{X}}_{2,c1}^{R}, ... \hat{\mathbf{X}}_{n,c1}^{R}\}$ & augmented training set of class 1\\
            
            ${\mathbf{X}}_{c2}^{tr} = \{ {{\mathbf{X}}}_{1,c2}^{CBT}, \hat{\mathbf{X}}_{1,c2}^{R}, \hat{\mathbf{X}}_{2,c2}^{R}, ... \hat{\mathbf{X}}_{n,c2}^{R}\}$ & augmented training set of class 2\\

			\bottomrule
		\end{tabular}
		
	\end{scriptsize}
\end{table}
\renewcommand{\arraystretch}{1.0}

\textbf{Augmentator network.}
As illustrated in Figure~\ref{fig:main}-A, augmentator network consists of a GAN architecture as a generative model with two GNN based components, namely the generator $G$ and the discriminator $D$. The two components compete against each other on training process to so that the architecture outputs synthetic data that is more congenerous with the patterns and variabilities in the training data. To provide this adversarial learning approach, the generator learns to imitate the training samples more accurately whereas the discriminator learns to discriminate the samples generated by the generator in a better way. For this purpose, we utilize the gGAN\citep{gGAN} as the GAN architecture. In the paper which gGAN was proposed, it was used for normalization-based  mapping with respect to a fixed population template. For our pipeline, we feed the CBT both as a fixed-population template and a training sample to use the gGAN as an augmentator. To the best of our knowledge, this is the first time that a single population template representing edge connectivity was used on a graph-based generative architecture. 

In order to perform class-based graph augmentation, we firstly train two different gGAN architectures independently for the existing two graph classes. For a given CBT denoted with ${\mathbf{X}}_{1}^{CBT}$, we give the CBT to gGAN's training sample input and fixed population template input. With that input provided, the gGAN optimized a loss function composed of adversarial loss and the L1 loss. The adversarial loss is defined as: 

\begin{gather}
    \mathcal{L}_{adv} = \mathbb{E}_{x\sim{\mathbf{X}}_{1}^{CBT}}[log(D(x))] + \mathbb{E}_{{\hat{x}}\sim{\mathbf{X}}_{1}^{CBT}}[log(1-D(G(\hat{x})))]
\end{gather}

The provide better robustness to the generated samples, gGAN also presents an additional L1 loss term controlled by a $\lambda$ parameter to its adversarial loss. The introduced L1 loss aims to minimize the distance between the ground truth CBT ${\mathbf{X}}_{1}^{CBT}$ and the generated synthetic graph $\hat{\mathbf{X}}_{1}^{CBT}$. The complete loss function of the network is defined as:

\begin{gather}
	\mathcal{L}_{comp} = \mathcal{L}_{adv} + \lambda \mathcal{L}_{L1}(G)
\end{gather}

\textbf{Generator.}
The generator network of gGAN consists of three layers of graph convolutional neural networks (GCNs). All convolutional layers of the generator utilize the dynamic edge-conditioned filter proposed in \citep{Simonovsky:2017}. Moreover, batch normalization \citep{batchnorm} and dropout \citep{dropout} layers are included after each of the dynamic edge-convolutional layers to provide faster convergence along with less likelihood of overfitting. With this proposed design, generator learns to output synthetic graph data $\hat{\mathbf{X}}_{1}^{CBT}$ by only examining one single population template ${\mathbf{X}}_{1}^{CBT}$ and learning from the non-Euclidean geometry of the graph utilizing edge-based convolution.  

\textbf{Discriminator.}
The discriminator network is composed of 2 graph convolutional layers which are exactly the same with layer used in the generator \citep{Simonovsky:2017} in order to discriminate between the synthetic brain graph $\hat{\mathbf{X}}_{1}^{CBT}$ outputted by the generator and the ground truth CBT ${\mathbf{X}}_{1}^{CBT}$. As in the generator network, all convolutional networks are followed by batch normalization and dropout. Upon each forward propagation, the discriminator network outputs a discriminator loss value in a different scale that evaluates how indistinguishable the ground truth CBT ${\mathbf{X}}_{1}^{CBT}$ and synthetic brain graph $\hat{\mathbf{X}}_{1}^{CBT}$ are. This way, gGAN architecture is designed in such a way that it maximizes the output of the discriminator loss while minimizing the generator loss which forms the basis for adversarial learning paradigm.

\textbf{Brain graph generation.}
As depicted in Figure~\ref{fig:main}-B, a set of symmetric random noise matrices of desired size $\{{{X}}_{1}^{R}, {{X}}_{2}^{R}, {{X}}_{3}^{R}, ..., {{X}}_{n}^{R} \}$ are passed through the trained generator to generate synthetic brain graphs. Since the real data represents edge connectivity of ROIs inside the brain, the random noise matrices with size $n_r \times n_r$ are also created in a symmetric manner. To have more variability in generated samples, random noise matrices are created on-demand during graph generation phase. Upon passing random noises through the independently trained generators, we extend the training sets of classes to ${\mathbf{X}}_{c1}^{tr} = \{ {{\mathbf{X}}}_{1,c1}^{CBT}, \hat{\mathbf{X}}_{1,c1}^{R}, \hat{\mathbf{X}}_{2,c1}^{R}, ... \hat{\mathbf{X}}_{n,c1}^{R}\}$ and  ${\mathbf{X}}_{c2}^{tr} = \{ {{\mathbf{X}}}_{1,c2}^{CBT}, \hat{\mathbf{X}}_{1,c2}^{R}, \hat{\mathbf{X}}_{2,c2}^{R}, ... \hat{\mathbf{X}}_{n,c2}^{R}\}$ rather than one-shot learning only on the CBTs.

\textbf{Graph classification.} 
To classify the two classes of brain graphs, we used a machine learning (ML) classifier rather than a deep graph classifier network due to data scarcity issue in our domain and dataset. Since each instance of the training set represents edge connectivity, each sample is symmetric and upper triangular part of the data is sufficient to define features to an ML classifier. Thereby, training sets of both classes are vectorized only using their upper triangular part which yields a vector of size $\frac{(n_r\times n_r) - n_r}{2}$ for each sample. Finally, we train a random forest classifier with the augmented dataset.

\section{Results}

\textbf{Clinical Dataset.}
In evaluation phase, we have used a dataset with 70 subjects (35 AD and 35 LMCI) from the Alzheimer's Disease Neuroimaging Initiative (ADNI) database GO public dataset \citep{Mueller:2005}. Each of the subjects consists of 6 morphological networks. The networks are derived from maximum principal curvature, the mean cortical thickness, the man sulcal depth, the average curvature measurements, cortical surface area, and the minimum principle area for both of the hemispheres. The cortical surface is reconstructed from T1-weighted MRI using the FreeSurfer \citep{Fischl:2012} and each view is defined with 35 ROIs by Desikan-Killiany atlas \citep{Desikan:2006}. In each of the views, the weight of the entry denote the strength of connectivity between given ROIs inside the brain network.

\textbf{Data Preprocessing.}
In order to examine the generalizability of our solution on population templates coming from different distributions, we have adapted 5-fold cross-validation method in a balanced fashion across the two classes. With the training folds, we have generated the CBTs for one-shot learning using the DGN (Deep Graph Normalizer) proposed in \citep{Gurbuz:2020} since DGN is one of the recent brain network fusion methods that is also capable of fusing single view brain graphs. Moreover, we have randomly sampled our dataset 20 time for each run of the 5-fold cross validation. Thereby, we report the performance measures as the average of 100 different folds that involves a population template coming from a different distribution.

\textbf{Hyperparameter Tuning.}
For the parameters of gGAN architecture, we have firstly set the hyperparameter of L1 loss coefficient $\lambda$ to 1000. Both for the generator $D$ and discriminator $D$, we use the ADAM optimizer. To optimize the loss of the generator ${G}$, we have set the learning rate to $0.5$ with $\beta_1 = 0.5$ and $\beta_2 = 0.999$. Finally, we set the parameters of the discriminator ${D}$ as learning rate of $0.1$ along with  $\beta_1 = 0.1$ and $\beta_2 = 0.999$. For each fold, we train the gGAN with $100$ epochs. Moreover, all the parameters are chosen by experimenting.

\begin{figure}[H]
     \centering
     \begin{subfigure}[b]{0.45\textwidth}
         \centering
         \includegraphics[width=\textwidth]{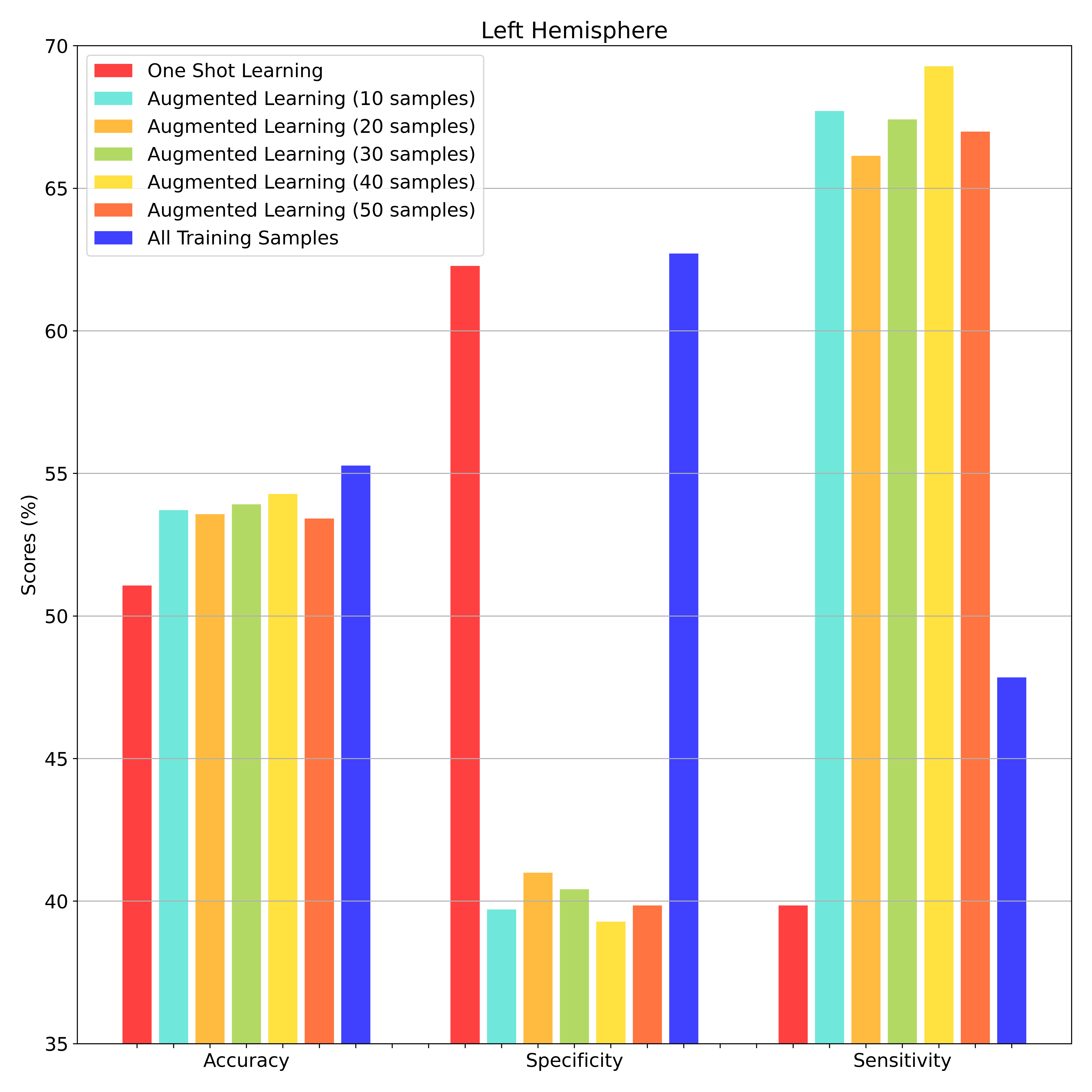}
         \caption{LH Augmentation Seperated}
         \label{fig:lh_detailed}
     \end{subfigure}
     \hfill
     \begin{subfigure}[b]{0.45\textwidth}
         \centering
         \includegraphics[width=\textwidth]{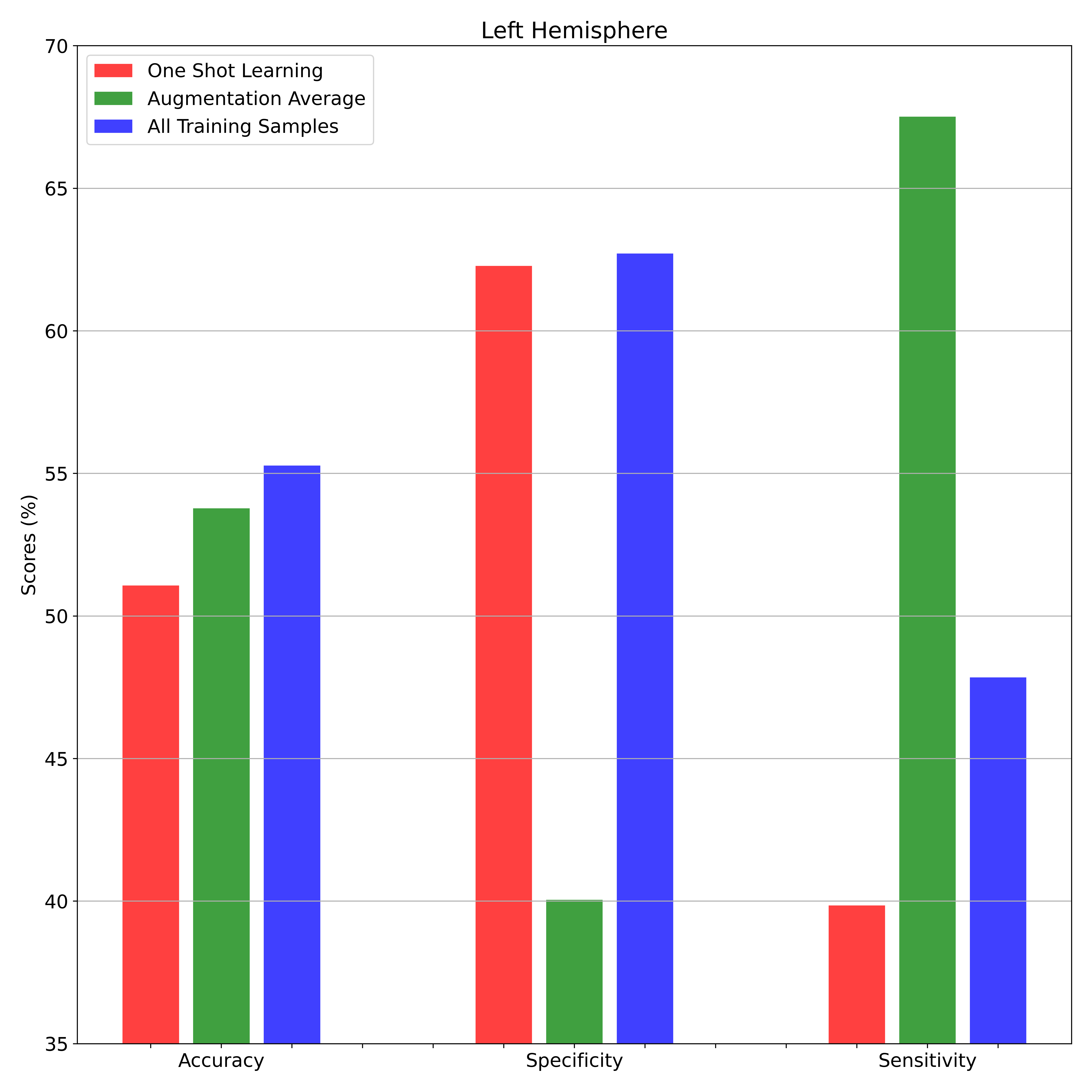}
         \caption{LH Augmentation Averaged}
         \label{fig:lh_summary}
     \end{subfigure}
     \hfill
        \caption{\emph{Benchmarks on classification (AD/LMCI) performance of augmented learning for left hemisphere.} Comparison of the one-shot learning with augmented learning with different number of augmented samples. We define the metrics of training with all samples as an upper bound and average across 100 folds.}
        \label{fig:lh_boost}
\end{figure}

\begin{figure}[H]
     \centering
     \begin{subfigure}[b]{0.45\textwidth}
         \centering
         \includegraphics[width=\textwidth]{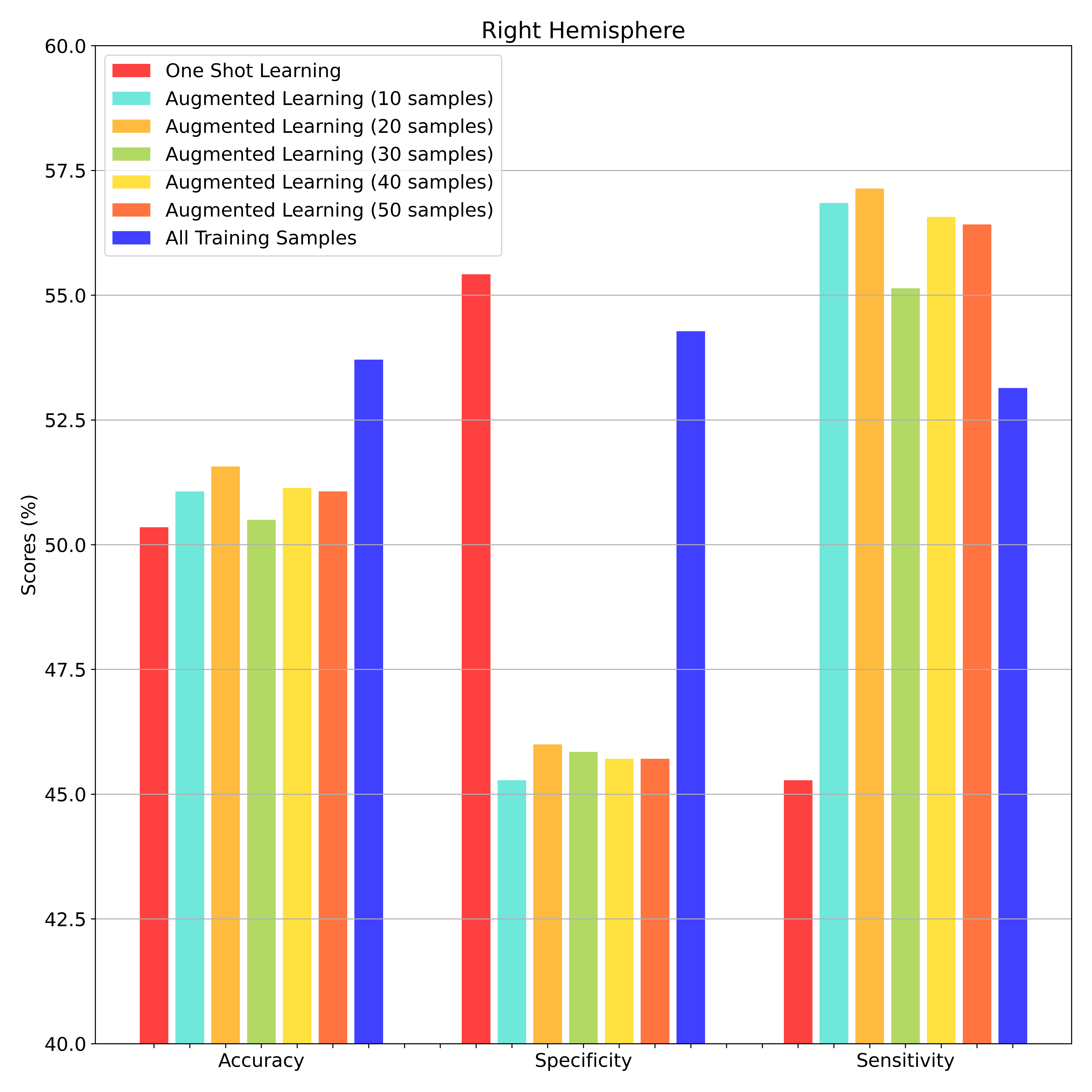}
         \caption{RH Augmentation Seperated}
         \label{fig:rh_detailed}
     \end{subfigure}
     \hfill
     \begin{subfigure}[b]{0.45\textwidth}
         \centering
         \includegraphics[width=\textwidth]{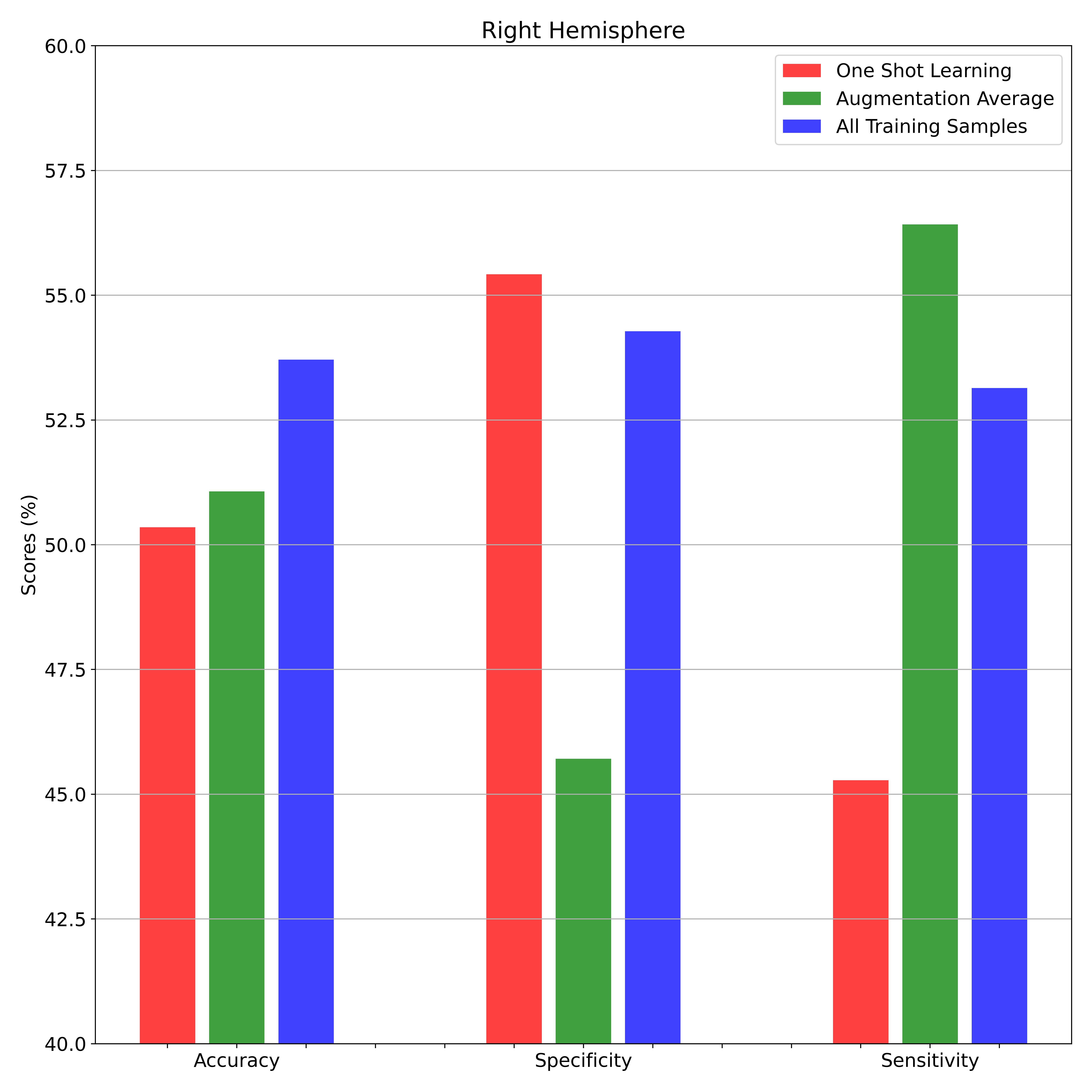}
         \caption{RH Augmentation Averaged}
         \label{fig:rh_summary}
     \end{subfigure}
     \hfill
        \caption{\emph{Benchmarks on classification (AD/LMCI) performance of augmented learning for right hemisphere.} Comparison of the one-shot learning with augmented learning with different number of augmented samples. We define the metrics of training with all samples as an upper bound and average across 100 folds.}
        \label{fig:rh_boost}
\end{figure}

\begin{figure}[H]
     \centering
     \begin{subfigure}[b]{0.4\textwidth}
         \centering
         \includegraphics[width=\textwidth]{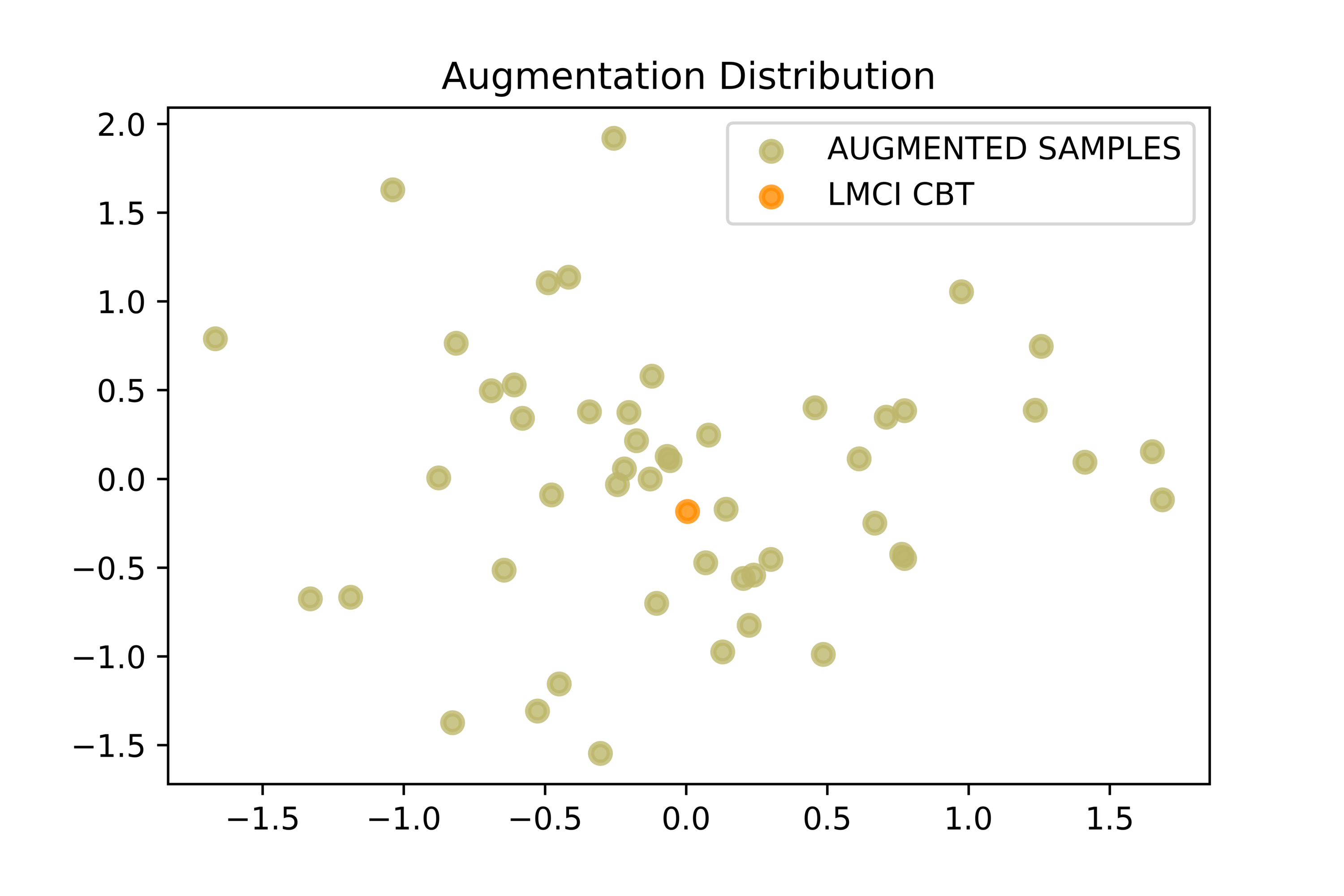}
         \caption{LH LMCI Augmentation}
         \label{fig:lh_ci}
     \end{subfigure}
     \hfill
     \begin{subfigure}[b]{0.4\textwidth}
         \centering
         \includegraphics[width=\textwidth]{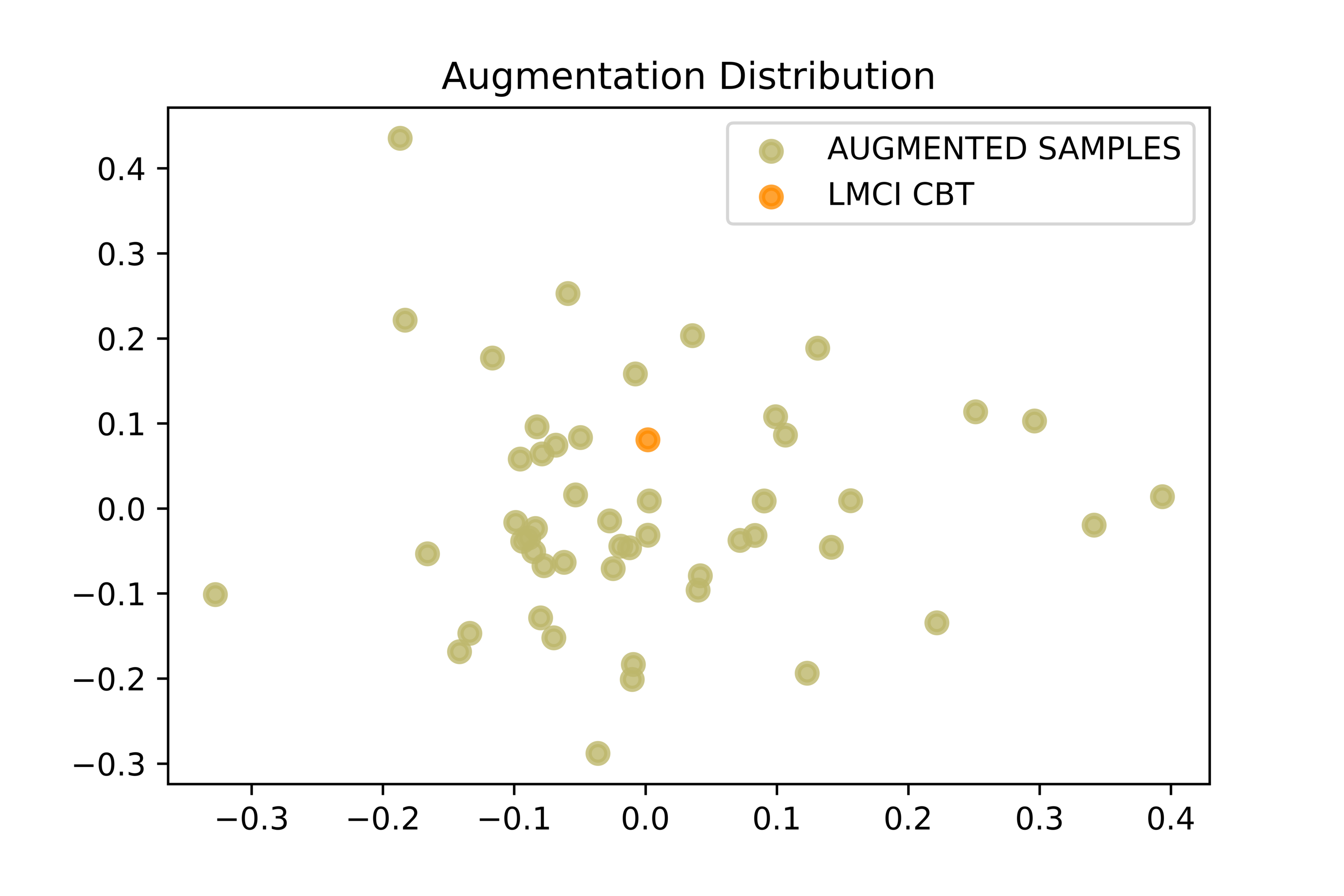}
         \caption{RH LMCI Augmentation}
         \label{fig:rh_ci}
     \end{subfigure}
     \hfill
     \begin{subfigure}[b]{0.4\textwidth}
         \centering
         \includegraphics[width=\textwidth]{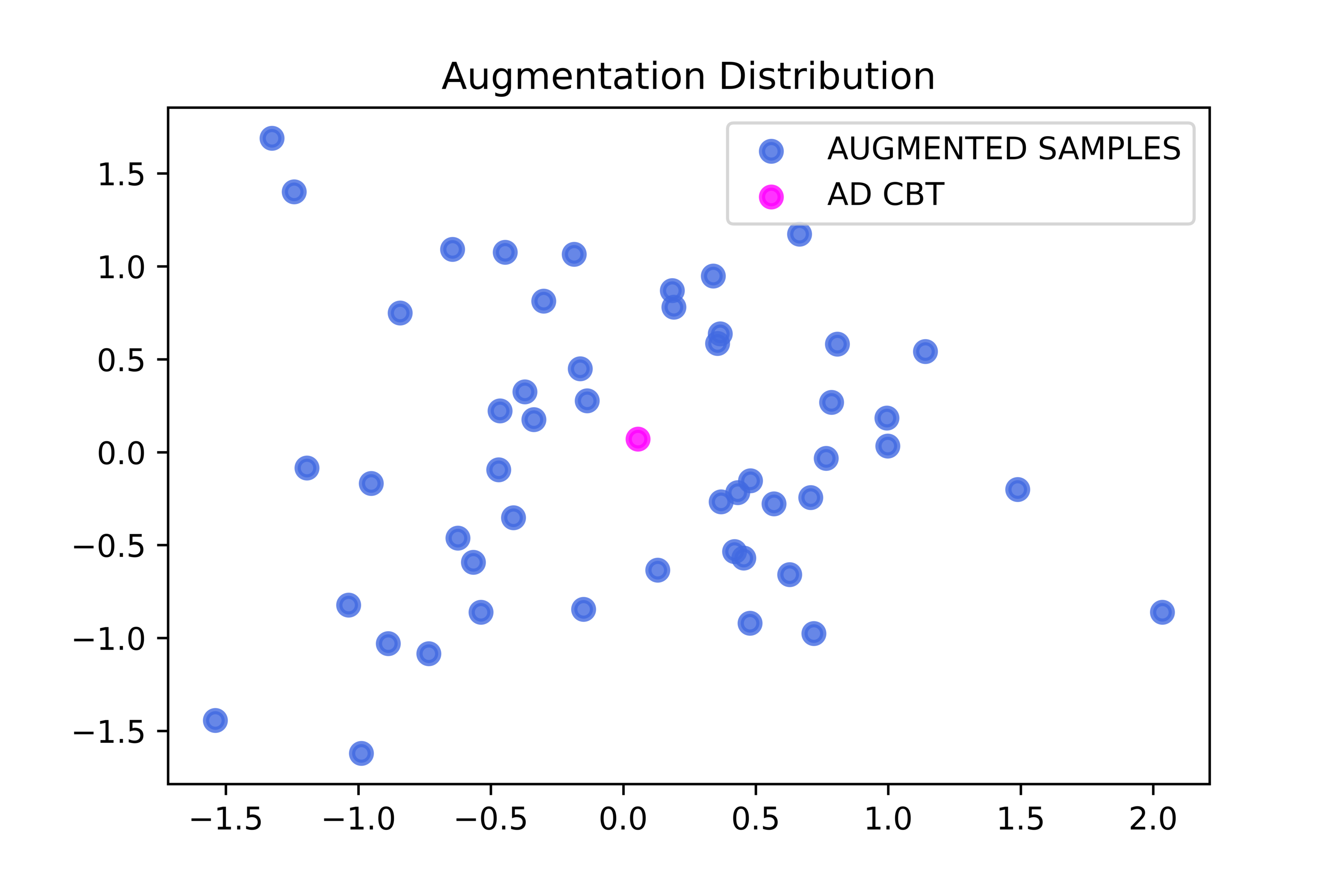}
         \caption{LH AD Augmentation}
         \label{fig:lh_ad}
     \end{subfigure}
     \hfill
     \begin{subfigure}[b]{0.4\textwidth}
         \centering
         \includegraphics[width=\textwidth]{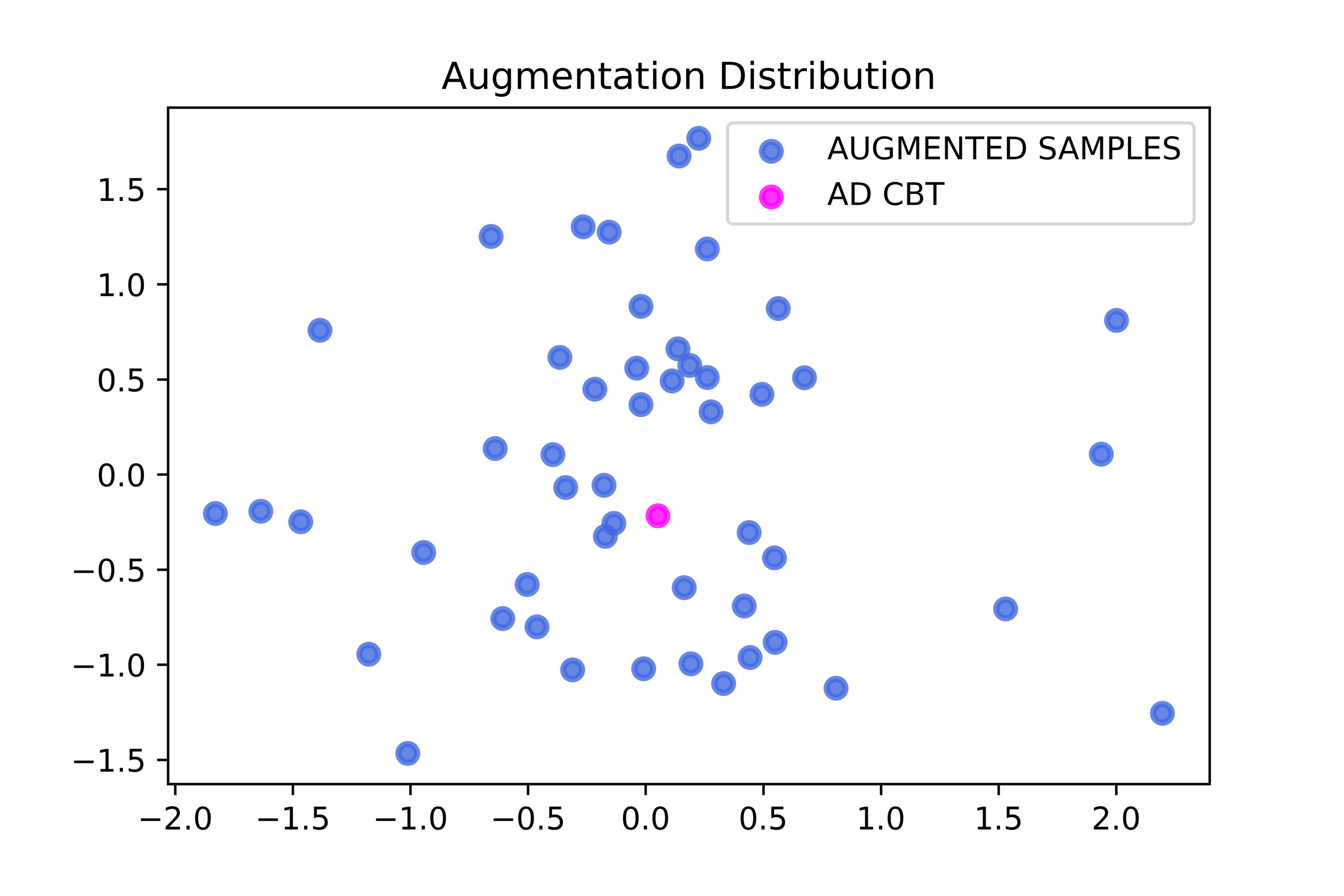}
         \caption{RH AD Augmentation}
         \label{fig:rh_ad}
     \end{subfigure}
     \hfill
        \caption{\emph{Distribution of the augmented samples.} PCA (Principal Component Analysis) plots of the augmented samples and the population template fitted in a 2D space for both hemispheres of AD and LMCI population.}
        \label{fig:pca}
\end{figure}

\begin{figure}[H]
	\centering
	\vspace{-10pt}
	{\includegraphics[width=\textwidth]{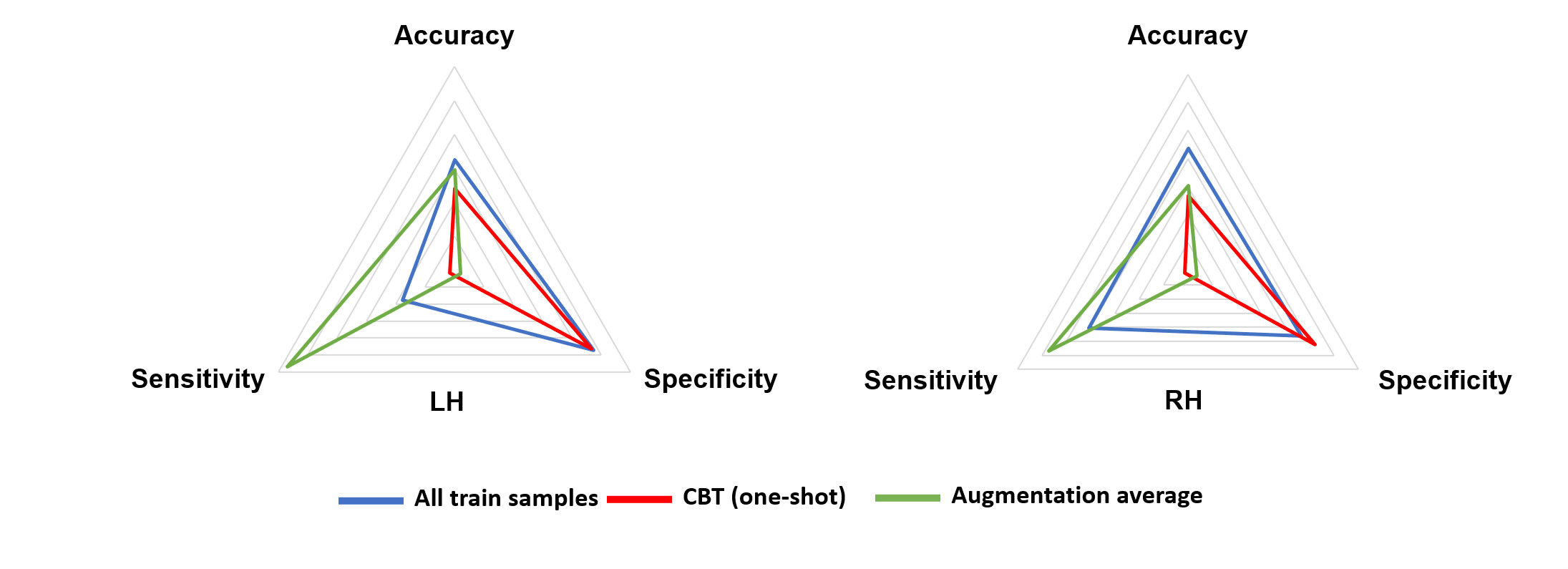}}
	\hfill
        \caption{\emph{Metric trade-offs.} Variances of metrics among different learning methods for both of the hemispheres.}
    \label{fig:metrics tradeoff}
\end{figure}

\textbf{Evaluation.}
To evaluate our augmentation network performance measures on a classification problem, we have compared the one-shot learning performance with the augmented learning promoted by our proposed network for both of the hemispheres and provided the results in Figure~\ref{fig:lh_boost} and Figure~\ref{fig:rh_boost}. The detailed metrics are also provided numerically in Table~\ref{tab:lh_table} and Table~\ref{tab:rh_table}. Along with the classification performance metrics, Figure~\ref{fig:metrics tradeoff} describes the generic trade-off trend on our augmentation technique for both hemispheres. Since the methods we have inspected require multiple training data for augmentation or designed specifically for imbalanced classes with multi-view graphs \citep{Graa:2019}, we have benchmarked only against the baseline of one-shot learning with CBT and the upper bound of training with all samples. Figure~\ref{fig:pca} illustrates the distribution of the generated data compared to the population template. To generate the plots, we have vectorized the upper-triangular part of the brain graphs and performed principal component analysis (PCA) on two dimensions. To present how efficient the generated samples capture the morphological connectional features of the population, we have provided circular brain network graphs of randomly taken subjects and marked top-$k$ connectivities for each population and hemisphere. For all metrics reported, the first view of the dataset is used which is the maximum principal curvature.

\begin{table}[H]

\begin{tabular}{c|ccc}
\hline
\multicolumn{1}{c|}{\textit{Left Hemisphere}}    & \multicolumn{3}{c}{\textit{Metrics}}                                                                      \\ \hline
\textbf{Learning Method} & \multicolumn{1}{c|}{\textbf{Accuracy}} & \multicolumn{1}{c|}{\textbf{Specificity}} & \textbf{Sensitivity} \\ \hline
All train samples        & \multicolumn{1}{c|}{\textcolor{blue}{$\bf55.28 \pm 10.28$}}    & \multicolumn{1}{c|}{\textcolor{blue}{$\bf62.71 \pm 17.24$}}       & \textcolor{blue}{$\bf47.85 \pm 16.33$}       \\
CBT (one-shot)           & \multicolumn{1}{c|}{\textbf{$\bf51.07 \pm 7.78$}}    & \multicolumn{1}{c|}{\textbf{{$\bf62.28 \pm 39.56$}}}       & \textbf{$39.85 \pm 38.71$}       \\
Augmented Learning (10 Samples)      & \multicolumn{1}{c|}{$53.71 \pm 7.92$}             & \multicolumn{1}{c|}{$39.71 \pm 36.50$}                & \ul{$67.71 \pm 37.77$ }               \\
Augmented Learning (20 Samples)        & \multicolumn{1}{c|}{$53.57 \pm 8.77$}             & \multicolumn{1}{c|}{\ul{$41.00 \pm 36.95$}}                 & $66.14  \pm 36.84 $             \\
Augmented Learning (30 Samples)        & \multicolumn{1}{c|}{\ul{$53.92 \pm 8.88$}}             & \multicolumn{1}{c|}{$40.42 \pm 37.14$}                & $67.42 \pm 36.98$                \\
Augmented Learning (40 Samples)        & \multicolumn{1}{c|}{{$\bf 54.28 \pm 9.31$}}             & \multicolumn{1}{c|}{$39.28 \pm 36.27$}                & {$\bf69.28 \pm 37.10$}                \\
Augmented Learning (50 Samples)        & \multicolumn{1}{c|}{{{$53.42 \pm 8.35$}}}       & \multicolumn{1}{c|}{$39.85 \pm 37.15$}                & $66.99 \pm 37.39 $               \\
Augmentation Average     & \multicolumn{1}{c|}{\bfseries{$ 53.78 \pm 8.69$}}    & \multicolumn{1}{c|}{\textbf{$40.05 \pm 36.81$}}       & \textbf{$67.51 \pm 37.23$}       \\ \hline

\end{tabular}
\caption{Comparison between one-shot learning and augmented learning with different number of augmented samples. Each case is averaged across 100 folds. Augmentation average is averaged over all cases of augmented learning in the table. \textcolor{blue}{Blue}: Upper bound, \textbf{Bold}: Best, \ul{Underlined}: Second best \label{tab:lh_table}}
\end{table}

\begin{table}[H]

\begin{tabular}{c|ccc}
\hline
\multicolumn{1}{c|}{\textit{Right Hemisphere}}    & \multicolumn{3}{c}{\textit{Metrics}}                                                                      \\ \hline
\textbf{Learning Method} & \multicolumn{1}{c|}{\textbf{Accuracy}} & \multicolumn{1}{c|}{\textbf{Specificity}} & \textbf{Sensitivity} \\ \hline
All train samples        & \multicolumn{1}{c|}{\textcolor{blue}{$53.71 \pm 11.84$}}    & \multicolumn{1}{c|}{\textcolor{blue}{$54.28 \pm 20.60$}}       & \textcolor{blue}{$53.14 \pm 19.69$}       \\
CBT (one-shot)           & \multicolumn{1}{c|}{\textbf{$50.35 \pm 7.78$}}    & \multicolumn{1}{c|}{\textbf{{$\bf55.42 \pm 39.60$}}}       & \textbf{$45.28 \pm 41.65$}       \\
Augmented Learning (10 Samples)        & \multicolumn{1}{c|}{$51.07 \pm 6.87$}             & \multicolumn{1}{c|}{$45.28 \pm 40.91$}                & \ul{$56.85 \pm 42.28$ }               \\
Augmented Learning (20 Samples)        & \multicolumn{1}{c|}{$\bf 51.57 \pm 7.18$}             & \multicolumn{1}{c|}{\ul{$46.00 \pm 39.77$}}                 & {$\bf 57.14  \pm 41.79 $}             \\
Augmented Learning (30 Samples)        & \multicolumn{1}{c|}{$50.50 \pm 7.64$}             & \multicolumn{1}{c|}{$45.85 \pm 40.11$}                & $55.14 \pm 41.60$                \\
Augmented Learning (40 Samples)       & \multicolumn{1}{c|}{\ul{$51.14 \pm 6.90$}}             & \multicolumn{1}{c|}{$45.71 \pm 39.58$}                & {$56.57 \pm 41.59$}                \\
Augmented Learning (50 Samples)        & \multicolumn{1}{c|}{{{$51.07 \pm 7.51$}}}       & \multicolumn{1}{c|}{$45.71 \pm 40.40$}                & $56.42 \pm 41.71 $               \\
Augmentation Average     & \multicolumn{1}{c|}{\bfseries{$ 51.07 \pm 7.24$}}    & \multicolumn{1}{c|}{\textbf{$45.71 \pm 40.16$}}       & \textbf{$56.42 \pm 41.80$}       \\ \hline

\end{tabular}
\caption{Comparison between one-shot learning and augmented learning with different number of augmented samples. Each case is averaged across 100 folds. Augmentation average is averaged over all cases of augmented learning in the table. \textcolor{blue}{Blue}: Upper bound, \textbf{Bold}: Best, \ul{Underlined}: Second best \label{tab:rh_table}}
\end{table}

\begin{figure}[H]
	\centering
	\vspace{-60pt}
	{\includegraphics[width=13cm]{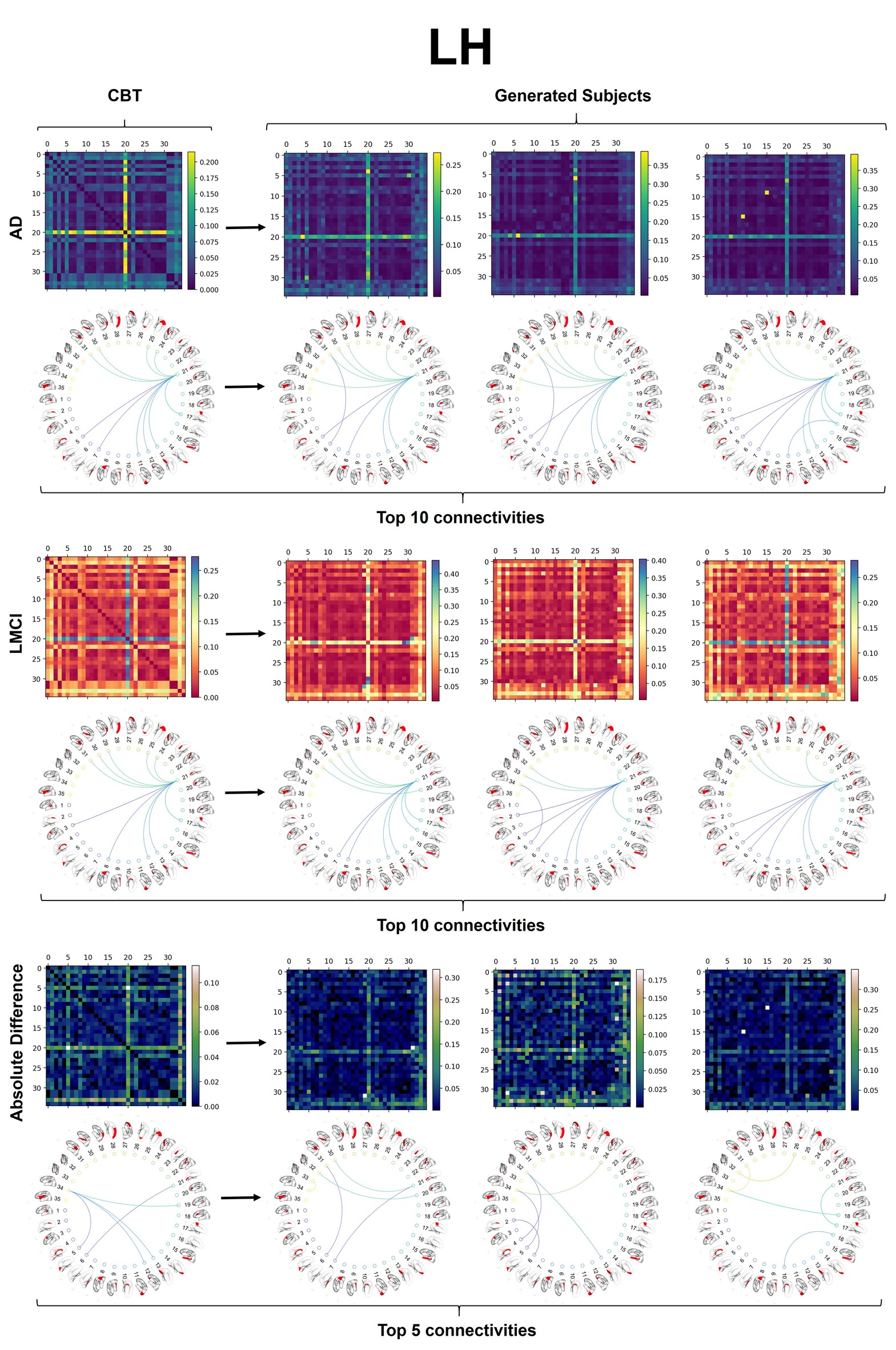}}
	\caption{\scriptsize \emph{Morphological connectional features of the CBT and synthetic brain graphs for left hemisphere.} Population template is compared with 3 randomly selected subject by maximal weights for both populations and absolute difference.} 
	\label{fig:lh_graph}
\end{figure}

\begin{figure}[H]
	\centering
	\vspace{-60pt}
	{\includegraphics[width=13cm]{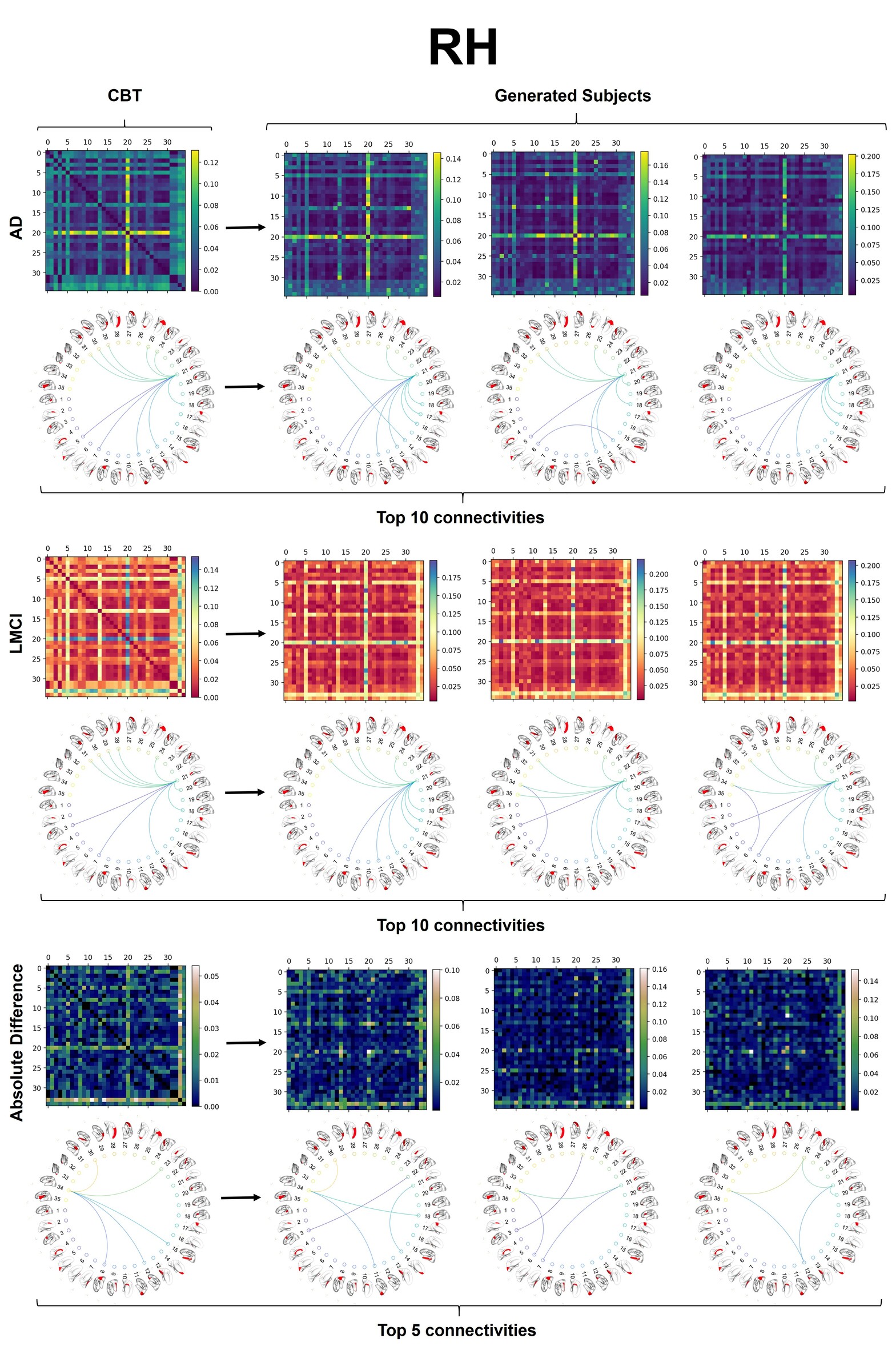}}
	\caption{\scriptsize \emph{Morphological connectional features of the CBT and synthetic brain graphs for right hemisphere.} Population template is compared with 3 randomly selected subject by maximal weights for both populations and absolute difference.} 
	\label{fig:rh_graph}
\end{figure}

\section{Discussion}
In this paper, we proposed an augmentator network originating from a single brain graph population template representing edge connectivity between brain regions. To the best of our knowledge, the proposed architecture investigates the idea of graph-based data augmentation from a single CBT in the context of network neuroscience for the first time. In particular, (i) we take on the compelling problem of performing augmentation using a single sample for clinical datasets suffering from data scarcity and (ii) query the augmentation and learning methods from a single CBT in terms of performance boost, variability and ability to capture population patterns by considering an uncommon approach.

\textbf{Classification performance boost.}
In evaluation phase of our augmentation network, we have achieved higher accuracy with the augmented learning compared to the one-shot learning performance. For both of the hemispheres, we have achieved closer results to our defined upper bound on average. Despite the consistent boost on both hemispheres with different number of augmented samples, we have observed a larger accuracy boost on left hemisphere with AD/LMCI classification. Our results on left hemisphere being more discriminative for AD/LMCI classification also coincides with previous studies as they yield higher overlap rates on discriminative bio-markers \citep{Gurbuz:2020}. As denoted in metric trade-off plots, we have also obtained a consistent trade-off of increasing sensitivity with both hemispheres on augmented learning. This trade-off can be interpreted as the augmented learning approach is better at distinguishing disordered brains. Even though our method can be utilized to any type of data representing edge connectivity provided a population template structure, we have inspected our method from a neuroscience perspective. With the consistent results we have obtained on a scarce AD/LMCI clinical dataset, we have shown that augmenting from a population template for a one-shot learning task may a reliable method to boost performance.

\textbf{Morphological connectional feature patterns of generated brain graphs.} In our study, we do not only focus on the performance boost of the neurological disorder classification problem but also examine the variability and morphological connectivity patterns of the synthetic brain graphs. In our experiments, our method yields a distribution of generated samples as if each sample was an authentic individual of the population regardless of the disorder and the hemisphere. This supports our idea of generating synthetic brains that are not biased by a single sample as opposed to conventional approaches. 
All these results show our proposed method's ability to generate samples that adapt the population specific patterns robustly. However, introducing an additional attention mechanism to track discriminative bio-markers across the brain regions can further enhance the classification step even more. For example, training the augmentator networks in a related manner by adding an additional loss component can generate synthetic brain graphs that are more discriminative between the populations. This way, enhanced model can be utilized to investigate the specific population pairs with provided brain graph templates.

\textbf{Reproducibility of discriminative bio-markers}.
In recent studies, \citep{Nebli:2022} proposed RG-Select to assess the reproducibility of discriminative bio-markers on different types of GNNs.  Motivated by this study, we have also analyzed the reproducibility of discriminative bio-markers distinguishing AD from LMCI  during the classification phase of our method with synthetic samples by randomly selecting generated samples for both classes and comparing with the population templates.  
Regarding the most relevant morphological connectivities including both hemispheres, AD and amnestic LMCI had in common the following most dissimilar morphological connections: (1) pericalcarine cortex (PrC) (region 21) with the precuneus (region 25), (2) PrC(region 21) with the superior temporal gyrus (region 30), (3) PrC with the lingual cortex (region 13) and (4) PrC with the latero-orbital frontal cortex (region 12). Rooting our findings in the state of the art, the implication of the precuneus, for an instance, has been demonstrated as a biomarker of early AD with a high functional connectivity in the precuneus and the frontal pole in the earliest stages of the disease \citep{Diasquale:2016,Jessica:2012}. Thus, we may speculate that dissimilar morphological connectivity with the precuneus and functional connectivity findings are intertwined as well where such dissimilarity could be interpreted as a coping mechanism rebounding against the undergoing neuro-degenerative process  in AD and a proof of brain-plasticity supporting  the cognitive reserve theory enhancing connections with spared cortical regions by remodeling neuronal networks \citep{Yaakov:2006}. A delicate analysis of the top 10 connectivities within the RH comparing AD and LMCI, pinpoints that PrC and precuneus connectivity is specific to AD patients and could thus be considered as a discriminative biomarker of AD. Another potential key to profiling AD patients brains in comparison with LMCI, is the particular cortical morphological connectivily joining the PrC and the cuneus, a cortical region  adjacent to the precuneus  and located posteriorly. In fact, \citep{Wu:2021}  established that decrease in cortical thickness of the cuneus is a risk factor of developing AD which was further supported by the findings of the genetic study of genetically determined AD.
Adopting the concept of cortical morphological connectivity in our study exceeded the morphology cortical-region-based approach as previously adopted by \citep{Yang:19} comparing cortical thickness in AD versus amnestic MCI. For an instance, such study pointed out that regions like the PrC, the inferior parietal cortex and the caudal middle frontal gyrus (CMFG) are significantly more concerned by cortical thinning in AD compared to LMCI, which is predictable as an outcome since amnestic MCI might be a prodromal stage of AD with a slighter severity and extent of the ongoing neurodegenerative process. Whereas, our study highlighted the dissimilarity of the connections between the PrC and the inferior parietal cortex (region 8) and the PrC with the CMFG as fingerprints of LMCI rather than AD. As a matter a fact, among patients presenting with amnestic MCI, \citep{Grambite:2011} stated that cortical thinning of the CMFG is a marker of executive dysfucntion in MCI. Changes in morphology (cortical thickness) is intertwined with morphological and functional connectivity. 

 In our analyses, we have observed discriminative connections on the transverse temporal cortex region of the brain consistently for both of the hemispheres. 
 Addition to the clinical studies, ReMI-Net architecture which is devised to forecast the population template of brains at a future timepoint determined the pericalcarine cortex as the most effected ROI by the Alzheimer’s disease for both left and right hemispheres \citep{Demirbilek:2021}. Thuswise, we have shown that synthetic brain graphs by our approach do not only boost the classification performance but also inherit the discriminative bio-markers from the population template profoundly and coincide with recent findings on AD/LMCI related clinical studies.

\textbf{Limitations and future work.}
Despite our promising results, our approach has certain limitations. First, the proposed method generate synthetic brain graphs only consisting of a single view which means each edge connectivity represent one and only one feature of a brain network feature. In our future work, we will focus on generating multi-view synthetic brain graphs which is a much more complex problem than fusing the view together. For that enhancement, a similar approach to topology-aware graph GAN architecture (topoGAN) which is proposed to predict multigraph brain views from a single source graph might be adapted \citep{Bessadok_topogan:2021}. Second, since the generated brain graphs are derived only from a single template, it has a certain bound in terms of variability even though it is able to catch the patterns. To address this limitation, an anchoring property could be introduced. Finally, the proposed network does not have an attention mechanism designed specifically for a classification task which might result in generated samples that are challenging to discriminate between population pairs. We aim to address those limitations in our future studies by enhancing our framework.

\textbf{Contributions.}
Mainly, we have adapted a compelling idea of augmenting from a population template of fused brain networks as opposed to conventional data augmentation techniques. Our approach can enable the generated samples to inherit the patterns and bio-markers of very high concern rather than relying on methods biased by a predefined number of subjects. In addition to capturing morphological patterns with GNNs, training the architecture with a single sample would be more efficient in terms of computational cost. Due to those reasons, we presume that data augmentation from a template will be a prominent method to consider for scarce brain connectomic data to identify neurological disorders.

\section{Conclusion}
In this paper, we proposed a GAN powered augmentation method to improve classification metrics by generating single-view brain network data using a single population template. Our method consists of two main steps for data augmentation. Firstly, we train the gGAN architecture independently for both populations using their template. Second, we feed random symmetric noise that imitates edge connectivity to the trained generators to create synthetic data. Overall, our method produced closer results to the upper bound metrics of the problem and generated synthetic samples that are capable of capturing population's bio-markers evaluated on an AD/LMCI dataset suffering from data scarcity. In our future work, we aim to include an attention mechanism to capture discriminative bio-markers better and extend our solution to multi-view data as well. 

\section{Acknowledgements}

This project has been funded by the 2232 International Fellowship for Outstanding Researchers Program of TUBITAK (Project No:118C288). However, all scientific contributions made in this project are owned and approved solely by the authors.

\newpage
\bibliography{Bib_unified}
\bibliographystyle{model2-names}

\end{document}